\documentstyle[12pt,psfig]{article}
\begin{document}
\vspace{-2cm}

\hspace{10.5cm}
\parbox{4cm}
{MPG-VT-UR 169/98
}
\\
\begin{center}
{\Large \bf Quark deconfinement and meson properties at finite temperature}
\bigskip

{\large D.~Blaschke$^1$, Yu.L.~Kalinovsky$^2$, P.C.~Tandy$^3$}
\date{}
\smallskip

$^1${\it Fachbereich Physik, Universit\"at Rostock, D-18051 Rostock, 
Germany}\\

$^2${\it Laboratory for Computing Techniques and Automatization, 
Joint Institute for Nuclear Research, 141980 Dubna, Russian Federation}\\

$^3${\it Center for Nuclear Research, Department of Physics, 
Kent State University, Kent, OH 44242, U.S.A.} 
\end{center}

\begin{abstract}
A simple confining separable interaction Ansatz for the rainbow-ladder 
truncated QCD Dyson-Schwinger equations is used to study quark deconfinement
and meson states at finite temperature.  The model is fixed at \mbox{$T=0$}
to implement quark confinement while preserving the Goldstone mechanism for
the $\pi$.  Within the Matsubara formalism, a very slow  temperature 
dependence is found for the $\pi$ and $\rho$ meson masses $M_H(T)$ until
near the deconfinement temperature $T_c=143$ MeV.  Related to rapid
decrease of the dynamically-generated  quark mass function for $T>T_c$,
this model produces $\pi$ and $\rho$ masses that rise significantly and 
are better interpreted as spatial screening masses. 
The $T$-dependent screening mass defect \mbox{$\Delta M_H=$}
\mbox{$2\pi T-M_H(T)$} is compared to results of lattice gauge theory 
simulations and also to those of an infrared dominant analytic model. 
\end{abstract}

\section{Introduction}

The experimental search for the QCD deconfinement phase transition in 
ultrarelativistic heavy-ion collisions will enter a new stage when the 
relativistic heavy-ion collider (RHIC) at Brookhaven will provide data 
complementary to results from the CERN SPS \cite{qm97}.
It is desirable to have a continuum field-theoretical modeling of quark 
deconfinement and chiral restoration at finite temperature and density (or
chemical potential $\mu$) that can be extended also to hadronic 
observables in a rapid and transparent way.  Significant steps in this
direction have recently been taken through a continuum approach to 
QCD$_{T,\mu}$ based on the truncated Dyson-Schwinger equations (DSEs) within 
the Matsubara formalism~\cite{bbkr,brs,mrs}.
For a recent review  see~\cite{br}.   A most appealing 
feature of this approach to modeling nonperturbative QCD$_{T,\mu}$ is 
that dynamical chiral symmetry breaking {\it and} confinement
is embodied in the the model gluon 2-point function constrained by chiral
observables at \mbox{$T=\mu=0$} and no new parameters are needed for
extension to \mbox{$T,\mu >0$}.   Approximations introduced by a 
specific truncation scheme for the set of DSEs can be systematically relaxed.
However due partly to the discrete Matsubara modes, the finite $T,\mu$ 
extension of current realistic DSE models entails a complicated set of 
coupled integral equations.   In the separable model we study here, detailed
realism is sacrificed in the hope that a few dominant and essential features
may be captured in a simple and transparent format.   We simplifiy an
existing \mbox{$T=\mu=0$} confining separable interaction Ansatz~\cite{b+} 
by using a gaussian form factor.  A related gaussian separable model has
recently been explored for meson properties~\cite{PB98}. 

\section{Model Dyson-Schwinger equations at $T= 0$}

In a Feynman-like gauge where we take $D_{\mu\nu}=\delta_{\mu\nu}D(p-q)$
to be an effective interaction between quark colored vector currents, 
the rainbow approximation to the DSE for the quark propagator 
$S(p)=[i \rlap/p A(p) + B(p) + m_0]^{-1}$ yields in Euclidean metric
\begin{eqnarray}
B(p) &=& 
\frac{16}{3} \int \frac{d^4q}{(2\pi)^4} D(p-q) 
\frac{B(q)+m_0}{q^2A^2(q)+\left[ B(q)+m_0\right]^2} \,\,\, , \\
\left[ A(p)-1 \right]  p^2 &=& 
\frac{8}{3} \int \frac{d^4q}{(2\pi)^4} D(p-q) 
\frac{(p\cdot q) A(q)}{q^2A^2(q)+\left[ B(q)+m_0\right]^2} \,\,\, .
\end{eqnarray} 
We study a separable interaction given by~\cite{b+}
\begin{equation}
D(p-q) = D_0~ f_0(p^2)f_0(q^2) + D_1~ f_1(p^2)(p\cdot q)f_1(q^2)~,  
\label{model}
\end{equation}
where $D_0, D_1$ are strength parameters and the form factors, for 
simplicity, are here taken to be
$f_i(p^2) = \mbox{exp}(-p^2/\Lambda_i^2)$ with range parameters $\Lambda_i$.
It is easily verified that  if $D_0$ is non-zero, then 
\mbox{$B(p)=\Delta m~f_0(p^2)$}, and if $D_1$ is non-zero, then  
\mbox{$A(p)=1+\Delta a~f_1(p^2)$}.   The DSE then reduces to nonlinear
equations for the constants $\Delta m$ and $\Delta a$.
The form factors should be chosen to
simulate the $p^2$ dependence of $A(p)$ and $B(p)$ from a more realistic 
interaction.   We restrict our considerations here to the rank-1 case 
where $D_1=0$ and \mbox{$A(p)=1$}.  The parameters $D_0$, $\Lambda_0$ 
and $m_0$ are used to produce reasonable $\pi$ and $\omega$ properties
as well as to ensure the produced  $B(p)$ has a reasonable strength with
a range $\Lambda_0 \sim 0.6 \dots 0.8$ GeV  to be realistic~\cite{b+}.

If there are no solutions to $p^2A^2(p)+(B(p)+m_0)^2=0$
for real $p^2$ then the quarks are confined. If in the chiral limit 
($m_0=0$) there is a nontrivial solution for $B(p)$, then chiral symmetry 
is dynamical broken.  Both phenomena are present in the separable model.
In the chiral limit, the model is confining if $D_0$ is strong enough to 
make $\Delta m/\Lambda_0\ge1/\sqrt{2{\rm e}}$.  Thus for a typical range
$\Lambda_0$,  confinement will typically occur with 
$M(p\approx 0)\ge 300$ MeV. 

Mesons as $q \bar q$ bound states are described by the Bethe-Salpeter 
equation which in the ladder  approximation for the present approach is
\begin{equation}
- \lambda_H(P^2) \Gamma_H(p,P) 
= \frac{4}{3} \int \frac{d^4q}{(2\pi)^4}
\left\{ 
D(p-q) \gamma_\mu S(q+\frac{P}{2}) \Gamma_H(q,P)S(q-\frac{P}{2}) \gamma_\mu 
 \right\}~,
\label{bs}
\end{equation}
where $P$ is the total momentum and $H$ denotes  a particular hadron.  
The quantity $\lambda_H$ is an  eigenvalue and at the physical mass
$M_H$ it satisfies \mbox{$\lambda_H(P^2=-M_H^2)=$}$1$.
With the rank-1 separable interaction,
only the $\gamma_5$ and the $\gamma_5 \rlap/P$ covariants contribute
to the $\pi$~\cite{b+},  and here we retain only the dominant form
$\Gamma_\pi(p,P) = i \gamma_5 E_\pi (p,P)$.  For the vector meson, the 
only surviving form is \mbox{$\Gamma_{\rho \mu}(p,P) =$}
\mbox{$\gamma_\mu^T(P) E_\rho (p,P)$}, with $ \gamma_\mu^T(P)$ being the
projection of 
$\gamma_\mu$ transverse to $P$.   The separable solutions have the form 
$E_i (p,P)=f_0(p^2) C_i(P^2), ~~i=\pi, \rho~$, where the $C_i$ factor out
from Eq.~(\ref{bs}).  The eigenvalues $\lambda_H(P^2)$ are the only 
determined quantities, and for example,
\begin{equation}
\lambda_\pi(P^2) = 
\frac{16 D_0}{3} \int \frac{d^4q}{(2\pi)^4}f_0^2(q^2) 
\left[(q^2-\frac{P^2}{4}) \sigma_V^+ \sigma_V^- + \sigma_S^+ \sigma_S^- 
\right]~.
\label{lampi}
\end{equation}
The quark propagator amplitudes are defined by 
\mbox{$S(p)=-i\rlap/p\sigma_V(p^2) + $}\mbox{$\sigma_S(p^2)$}, and the $\pm$
superscripts indicate momentum arguments $q\pm P/2$ respectively.
For the transverse and longitudinal $\rho$ states, there are expressions 
for $\lambda_\rho^T(P^2)$ and $\lambda_\rho^L(P^2)$ analogous to 
Eq.~(\ref{lampi}). 

With parameters $m_0/\Lambda_0=0.0096$, $D_0 \Lambda_0^2=128$ and 
$\Lambda_0=0.687$ GeV, this model yields  $M_\pi=0.14$ GeV, 
$M_\rho=M_\omega=0.783$ GeV, $f_\pi=0.104$ GeV, a quark condensate 
$\langle \bar q q\rangle^{1/3}=-0.248$ GeV, and a $\rho-\gamma$ coupling 
constant $g_\rho=5.04$.

In the limit where a zero momentum range for the interaction is simulated 
by \mbox{$f_0^2(q^2)\rightarrow$}\mbox{$\propto \delta^4(q)$}, then the
expressions for the various BSE eigenvalues $\lambda_H(P^2)$ reduce to
those of the Munczek and Nemirovsky~\cite{mn} model which implements
extreme infrared dominance via   
\begin{equation}
D(p-q) = \frac{3}{16} (2\pi )^4 \eta^2 \delta^{(4)}(p-q)~. 
\end{equation} 
The correspondence is not complete because the quark DSE solution in this 
model has $A(p)\neq 1$. We use $\eta=1.107$ GeV to produce the same $M_\omega$.
The finite temperature and chemical potential generalization of this infrared
dominant (ID) model has been studied in \cite{brs,mrs}.
 
\section{Finite temperature extension}

The generalisation to $T\neq 0$ is systematically 
accomplished by transcribing the Euclidean quark momentum via
$q \rightarrow (\omega_n, \vec{q})$, where $\omega_n=(2n+1)\pi T$ are the
discrete Matsubara frequencies.  Thus the integration measure in previous 
expressions becomes
\begin{eqnarray}
\int \frac{d^4q}{(2\pi)^4} \longrightarrow 
T \sum_{n=-\infty}^{\infty} 
\int \frac{d^3q}{(2\pi)^3} ~.
\end{eqnarray}
At finite temperature, the $O(4)$ symmetry is broken and the employed
mass shell condition must be specified.  One expects that
if there is a bound $\bar q q$ meson state, then the associated pole 
contribution to the relevant propagator or correlator will have a 
denominator proportional to 
\begin{equation}
1-\lambda_H(\Omega_m^2, \vec{P}^2) \; \propto  \; 
\Omega_m^2 + \vec{P}^2 +M_H^2~.
\label{dennom}
\end{equation}
We investigate the meson mode eigenvalues $\lambda_H$ using only the
lowest meson Matsubara mode ($\Omega_m=0$) and the continuation
\mbox{$\vec{P}^2\longrightarrow -M_H^2$}.   The masses so identified are
spatial screening masses corresponding to a behavior $\exp(-M_H x)$ in the
conjugate 3-space coordinate $x$ and should correspond to the lowest bound
state if one exists.   

\begin{figure}[htb]
\centerline{
\psfig{figure=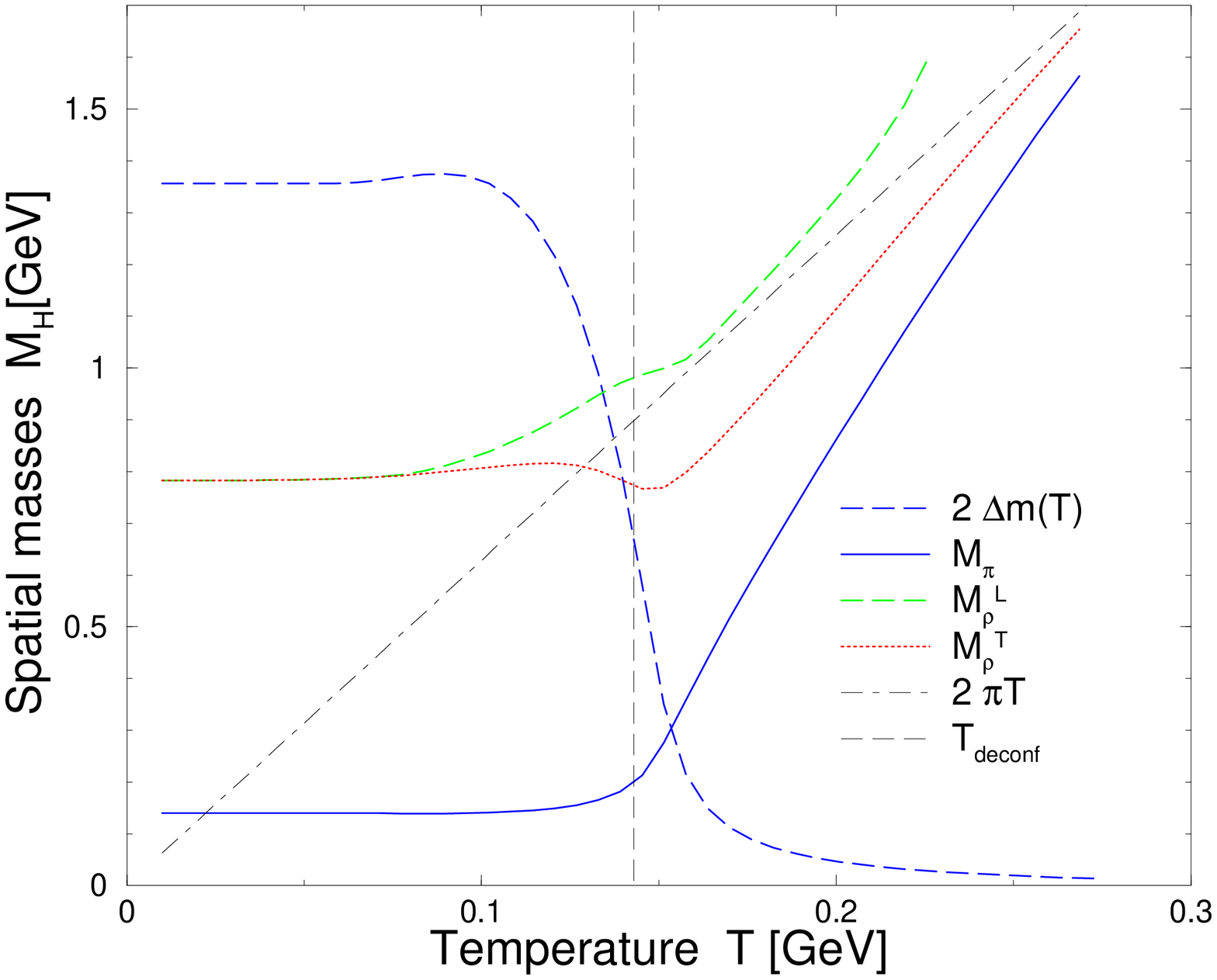,width=7cm,height=7cm,angle=-90}
\psfig{figure=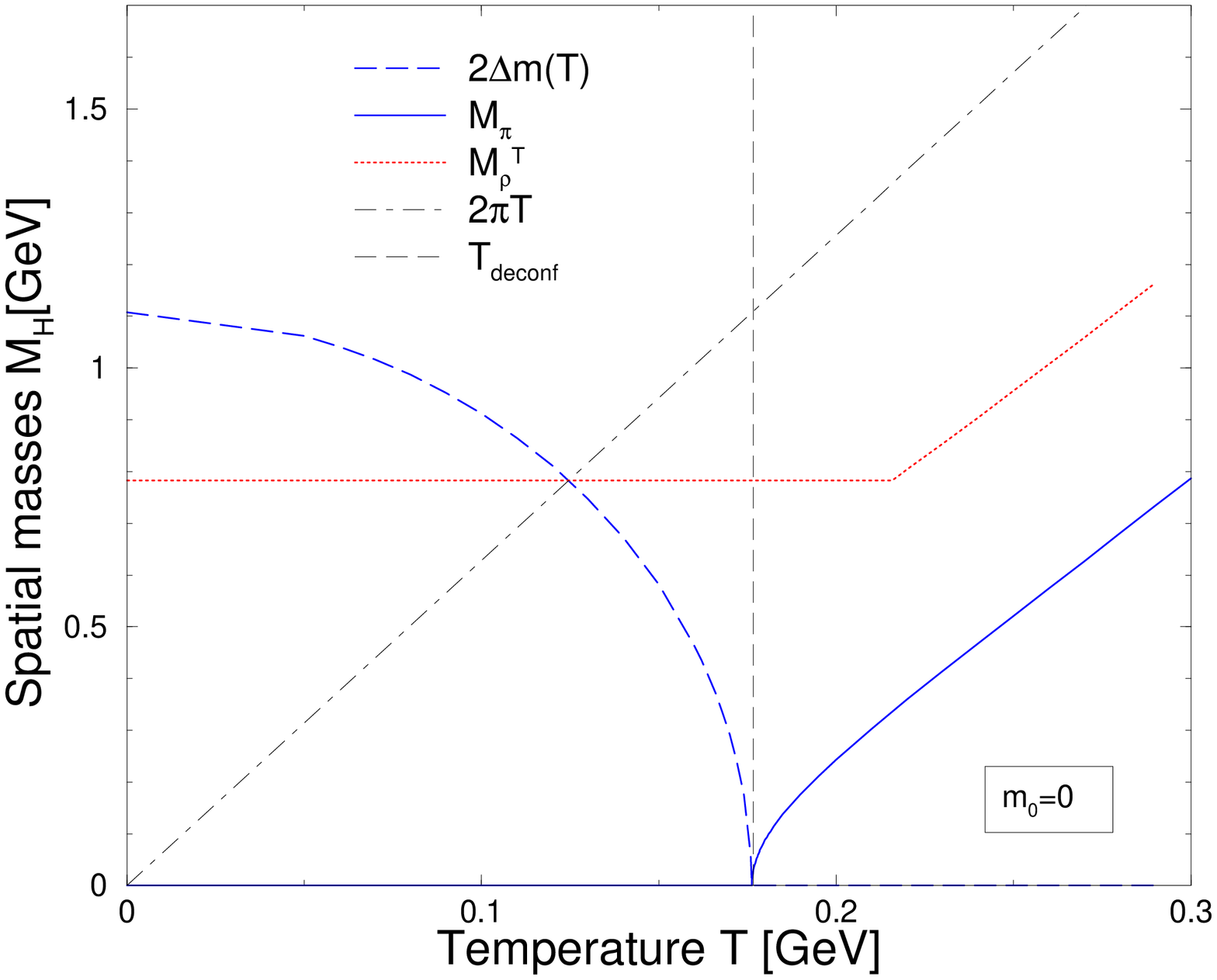,width=7cm,height=7cm,angle=-90}
}
\caption{Spatial meson masses of $\pi$ and $\rho$ for the Gaussian separable 
model (left panel) and for the analytic ID model (right panel).}
\end{figure}

The results are displayed in the left panel of Fig.~1 along with the 
strength ($\Delta m$) of the dynamical mass function at \mbox{$p=0$}.   The
deconfinement temperature is indicated.  The $\pi$ and $\rho$ masses are
only weakly dependent upon $T$ until near the transition.  For the $\pi$
this agrees with similar behavior obtained in the DSE approach with a more 
realistic interaction~\cite{bbkr}.  For both the 3-space transverse and 
3-space longitudinal $\rho$, the separate but still  weak temperature
dependence agrees with the findings~\cite{mrs} from the closely related 
ID model.  The right panel of Fig. 1 displays the $\pi$ and
transverse $\rho$ results from the ID model in the chiral limit.  
In both
panels of Fig. 1 we display the $T$-dependence obtained from the eigenmass
condition \mbox{$\lambda_H =1$} for $T>T_c$ in comparison with $2\pi T$.  

We interpret the model results rising similar to $2\pi T$
in the following way.   In the chiral limit and just above the transition, 
the dynamically generated mass function of the quarks has disappeared 
and the axial vector Ward identity that ties the pion Bethe-Salpeter
amplitude in the $\gamma_5$ channel to this quantity indicates that there
should be no solution of the pion BSE there.  However the 
particular model used here to estimate $\lambda_H$  assumes the existence 
of a bound state with an amplitude or wave function profile given by the 
chosen form factor $f_0(q^2)$.  The form of the latter was chosen at
$T=0$  to simulate the qualitative features of the DSE (and hence $\pi$
BSE) solution.  If bound state conditions disappear with rising 
temperature as interactions weaken, this approach is not well suited to 
detect it.
The high temperature form of the denominators of the quark propagators 
in the loop integral for the ``polarization'' function $\lambda_H$ is
\mbox{$\omega_n^2 + (\vec{q}\pm \vec{P}/2)^2$}.  The dependence of the 
form factor on $\omega_n^2$ emphasizes the lowest mode.   
For \mbox{$\omega_0 >> \Lambda_0$} the form factor dictates that non-zero 
relative momentum $q$ makes a minor contribution to the location 
of the lowest singularity in $\vec{P}^2$ that determines the spatial 
screening mass.  The higher the temperature the more the model behaves 
as if the fixed 3-space profile of the amplitudes had support 
effectively concentrated at $q=0$, that is, almost independent of relative
separation of quark and antiquark.   Thus an approach to a screening mass 
characteristic of an independent fermion pair is to be expected here. 
In Fig.~2 this point is emphasized via  comparison with spatial screening 
masses taken from~\cite{gocksch} and representing lattice simulations
above the transition.   The horizontal lines mark $2\pi$, which in the right
panel is corrected~\cite{gocksch} for the lattice time extent $N_t$.  

\begin{figure}[htb]
\centerline{
\psfig{figure=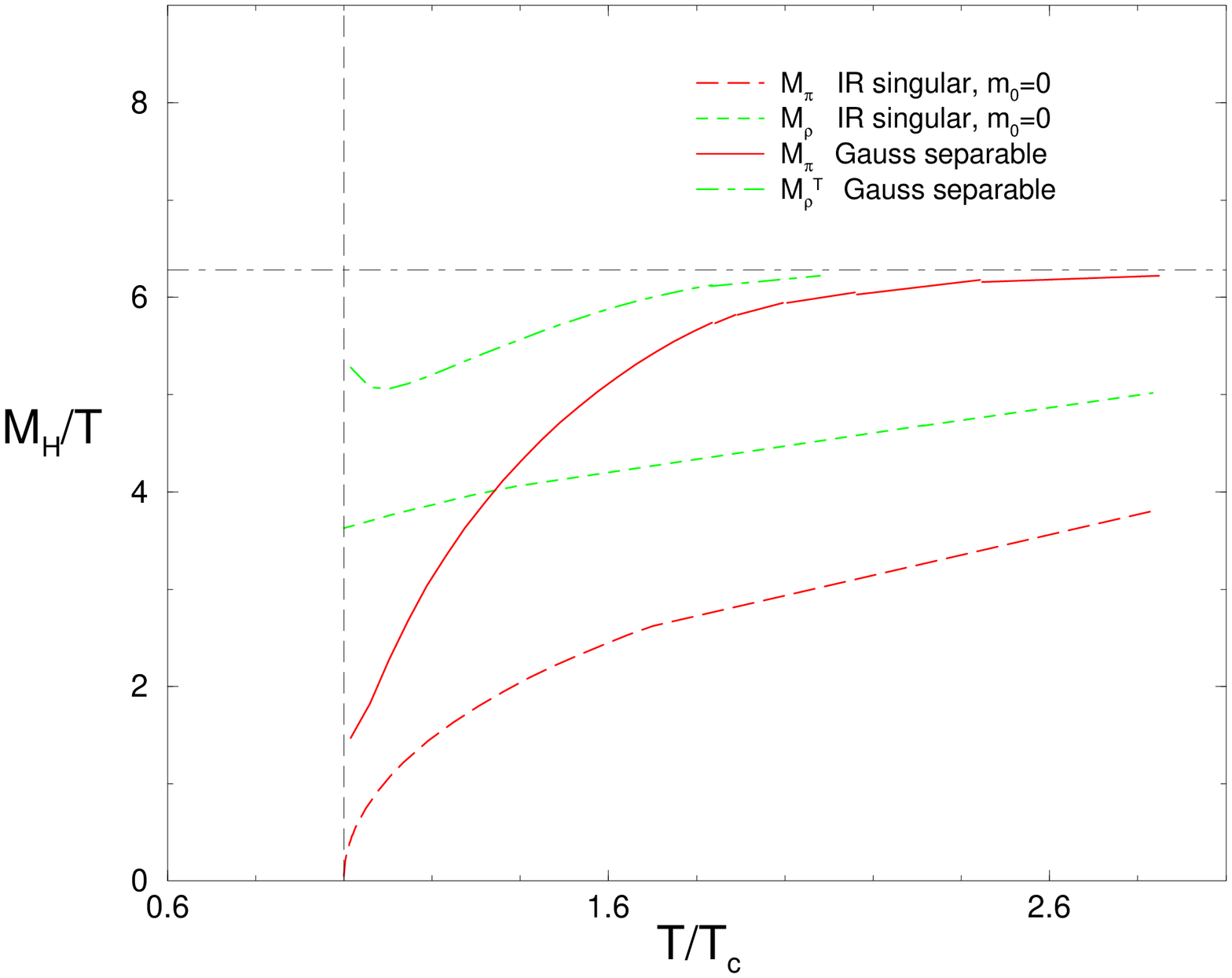,width=7cm,height=7cm,angle=-90}
\psfig{figure=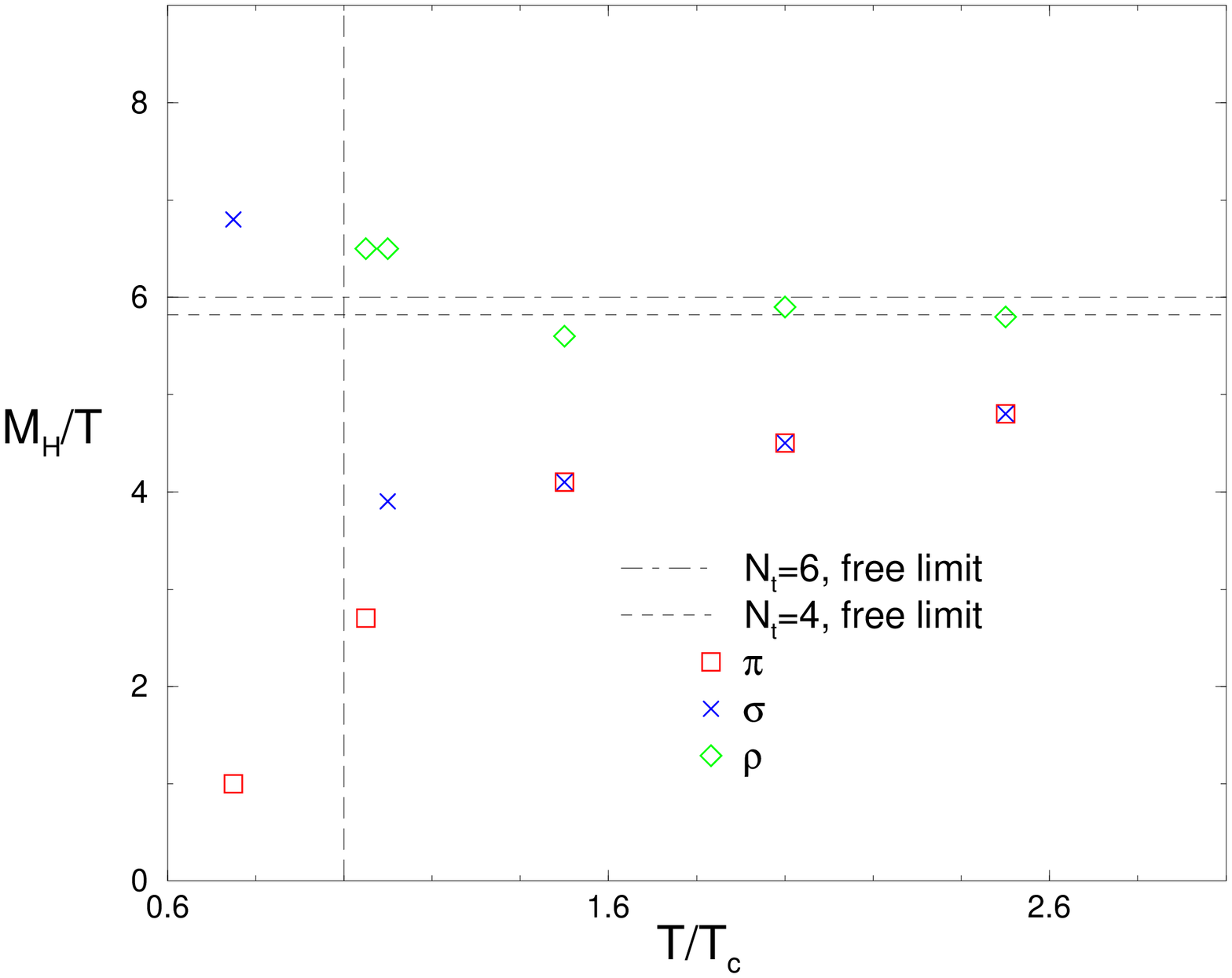,width=7cm,height=7cm,angle=-90}
}
\caption{Hadronic screening masses from the gaussian separable and ID 
models (left panel) compared to those of lattice gauge theory simulations 
(right panel, data taken from \protect \cite{gocksch})}
\end{figure}

From Figs.~1 and 2 it is evident that for \mbox{$T>T_c$}, the 
``screening mass defect'' \mbox{$\Delta M_H=$}\mbox{$2\pi T-M_H(T)$} 
is small and decreasing rapidly in the 
separable model, it is apparently larger in the lattice simulations 
especially for the $\pi$, and it is significantly stronger with a larger 
temperature range in the ID model.   The latter we attribute to the strong  
quark self-energy amplitude $A(p)>1$ in the ID model which is known to
significantly slow the approach to Stefan-Boltzmann 
thermodynamics~\cite{brs}.  Such effects should be investigated further 
with a finite range interaction and the rank-2 version~\cite{bkt} of 
the present separable model might help in that regard.

\section*{Acknowledgement}
Yu.L.K. thanks the {\sc Deutsche Forschungsgemeinschaft}
for financial support under Grant No. 436 RUS 17/125/97. 
P.C.T. acknowledges and the hospitality of the JINR Dubna and the
University of Rostock, where part of this work has been done with  support 
by the {\sc National Science Foundation} under Grant No. INT-9603385.
The research visits of D.B. at Kent State University and JINR Dubna
have been supported by {\sc Deutscher Akademischer Austauschdienst} and by
the Heisenberg-Landau programme, respectively.   Discussions with
E. Laermann, C. D. Roberts and S. Schmidt have been helpful in this work.

\end{document}